\documentstyle[multicol,aps,prl,epsf]{revtex}
\newcommand{\Vec}[1]{\mbox{\boldmath$#1$}}

\begin{document}

\draft

\title{
Hybridization-induced superconductivity from the electron 
repulsion on a tetramer lattice 
having a disconnected Fermi surface 
}

\author{
Takashi Kimura$^{1,2,3}$, Yuji Zenitani$^4$, 
Kazuhiko Kuroki$^5$, Ryotaro Arita$^1$, and Hideo Aoki$^1$
}

\address{
$^1$Department of Physics, University of Tokyo, Hongo, Tokyo 113-0033, Japan\\
$^2$Graduate School of Frontier Sciences, University of Tokyo, 
Hongo, Tokyo 113-0033, Japan\\ 
$^3$Advanced Research Institute for Science and Engineering,
Waseda University, Tokyo 169-8555, Japan \\
$^4$Department of Physics, Aoyama-Gakuin University, Chitosedai, Setagaya,
Tokyo 157-8572, Japan\\
$^5$ Department of Applied Physics and Chemistry, University of
Electro-Communications, Chofu, Tokyo 182-8585, Japan\\
}

\date{\today}

\maketitle

\begin{abstract}
Plaquette lattices with each unit cell containing 
multiple atoms are good candidates for disconnected 
Fermi surfaces, which are shown by Kuroki and Arita to 
be favorable for spin-flucutation mediated 
superconductivity from electron repulsion.  
Here we find an interesting example in a tetramer lattice where the structure 
within each unit cell dominates the nodal structure of the gap function.  
We trace its reason to the way in which a Cooper pair is formed 
across the hybridized molecular orbitals, 
where we still end up with a $T_c$ much higher than usual.
\end{abstract}

\medskip


\begin{multicols}{2}
\narrowtext 
The discovery of high-$T_c$ superconductors\cite{Bednorz} 
has kicked off renewed interests in 
electronic mechanisms for superconductivity.  
In the course of studies it is becoming increasingly 
clear that superconductivity can 
arise from repulsive interaction between electrons. 
The essence there is electrons interact 
by exchanging spin fluctuations, which can effectively 
act as an attraction in the gap equation, 
if we consider anisotropic pairing with 
nodes in the gap function. 

For the Hubbard model, 
this was first suggested from early calculations 
by Scalapino et al. \cite{Scalapino}. 
Subsequently a quantum Monte Carlo calculation \cite{Kuroki&Aoki} 
indeed indicated an enhancement of the pairing 
correlation with $d$-wave symmetry 
in the repulsive Hubbard model.   
The superconducting critical temperature ($T_c$) 
has been estimated with the fluctuation-exchange 
approximation (FLEX), a kind of renormalized 
random phase approximation.\cite{Bickers,Grabowski,Dahm} 

One remarkable point is $T_c \sim O(0.01t)$, 
esitmated for the repulsive Hubbard model in the 
two-dimensional (2D) square lattice, is 
{\it two orders of magnitudes smaller than} the 
starting electronic energy (i.e., the hopping integral $t$), 
although this gives the right order for the curates' $T_c$. 
Arita et al.  then looked at various lattices 
(square, trianglar, fcc, bcc, etc) in search of 
the most favorable case with a FLEX analysis\cite{AKA}. 
The best case, as far as these ordinary lattices are 
concerned, turns out to be the 2D square lattice, 
so $T_c < O(0.01t)$ remains.  
As discussed in Ref. \cite{Kuroki&Arita}, 
there are good reasons 
why $T_c$ is so low, where an important one is the 
presence of nodes in the superconducting gap function greatly 
reduces $T_c$:  While the main pair-scattering, 
across which the gap function has opposite signs to 
make the effective interaction attractive, 
some of the pair scatterings around the nodes 
have negative contributions to the effective attraction 
by connecting $k$-points on which the gap has the same sign. 

So a next important avenue to explore 
is: can 
we improve the situation by going over to multiband systems.  
Kuroki and Arita \cite{Kuroki&Arita} 
have shown that this is indeed the case if we 
have {\it disconnected Fermi surfaces}.  In this case 
$T_c$ is dramatically enhanced, because the sign change 
in the gap function can avoid the Fermi pockets, 
where all the pair-scattering processes contribute 
positively. \cite{Kuroki&Arita,NTT,KTA} 
This has been numerically shown to be the case for 
the triangular lattice (for spin-triplet pairing) 
\cite{KAfortri} and 
a squre lattice with a period-doubling \cite{Kuroki&Arita}, 
where $T_c$ as estimated with FLEX is as high as $O(0.1t)$.  

To be more precise, the key ingredinents are: 
(a) when the Fermi surface is nested, 
the spin susceptibility $\chi(\Vec{q},\omega)$ 
has a peak, with a width $\Delta \Vec{Q}$.  
(b) When a multiband system with a disconnected Fermi surface 
has an inter-pocket nesting 
(i.e., strong inter-pocket pair scattering 
and weak intra-pocket one)
the gap function has the same sign ($s$-wave symmetry) 
within each pocket, and the nodal lines can happily 
run in between the pockets.   

Here a natural question is: to prepare 
multiband systems systematically, 
we can consider lattices comprising some units, 
or ``plaquette lattices".  
This is indeed 
considered by Kimura et al. \cite{NTT} who conceived an idea of 
actually fabricating the structure from arrays of quatum dots. 
One example is the inset of Fig. 1, where the unit is a square 
lattice of squres, where the intra-plaquette transfer 
is stronger than the inter-plaquette one.  
Now, an important question is: how the structure of 
the superconducting gap function 
having higher $T_c$'s  is determined 
and whether the $s$-wave symmetry on each pocket is 
compulsory.   

In this paper, we study the repulsive
Hubbard model on a square lattice of diamonds, where 
a diamond is a four-site unit rotated by 45 degrees (Fig. 1). 
One motivation is that this structure is reminiscent of the 
in-plane lattice of B and C in CaB$_2$C$_2$ \cite{CaB2C2},
so should not be too unrealistic, although we are not 
claiming to consider this particular material.  
We have solved Eliashberg's equation \cite{Eliashberg} 
for the Hubbard model with the FLEX 
and obtained a $T_c$ that is considerably higher 
than that of the square lattice. 
More importantly, however, 
the present model, despite of its disconnected Fermi surface, 
has sign changes in the gap function within each pocket, 
so a different mechanism should be at work. 
We can trace back its reason in {\it real space} 
that the singlet pairing here results from a hybridization of two {\it molecular
orbitals} in each plaquette. 
\begin{figure}
\begin{center}
\leavevmode
\epsfxsize=6cm
\epsfysize=4cm
\epsfbox{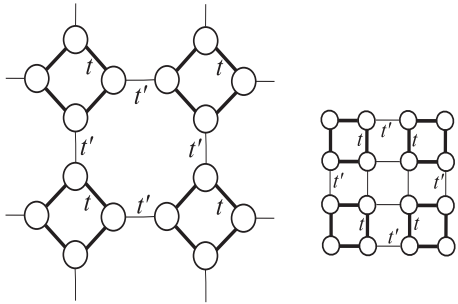}
\vspace*{0.5cm}
\caption{
The lattice considered in this paper.  
The inset shows the plaquette lattice in Ref. \protect\cite{NTT}.
}
\label{fig1}
\end{center}
\end{figure}

So a message of this work is that, 
in order to search for higher $T_c$'s, 
we should look not only at the Fermi surface 
but also at the atomic structure in real space.

We have used the four-band version of 
FLEX,\cite{NTT,KTA,Kurokietal} 
where the Green's function $G$, 
the spin susceptibility $\chi$, the self-energy $\Sigma$, and 
the superconducting gap function $\phi$ are $4\times 4$ matrices, 
e.g., $G_{lm}(\Vec{k}, i\varepsilon_n)$, where $l,m$ specify 
the four sites in a unit cell.  We can go from the site 
indeces over to band indeces with a unitary transformation. 
For the spin susceptibility, we concentrate on its 
largest eigenvalue denoted as $\chi$. 
From Green's function and the spin susceptibility 
we obtain the superconducting gap function 
(with spin-singlet pairing assumed) 
and $T_c$ by solving the linearized Eliashberg's equation\cite{Eliashberg}. 
In our analysis, we take $32\times 32$ $k$-point meshes and 
up to 4096 Matsubara frequencies, 
where the numerical results are sufficiently converged. 
The on-site Hubbard repulsion is taken to be a 
typical strong-correlation value, $U/W=7/6$, 
where $W$ is the band width.  
The band filling (=number of electrons/number of sites) 
is taken to be a value $n=0.85$, which is close, 
but not too close, to the half-filling where 
the Mott transition and antiferromagnetic order are 
expected.  In the below we have confirmed that 
when the eigevalue of Eliashberg's equation becomes unity 
prior to the divergence of the spin susceptibility.  

The FLEX result for the spin susceptibility, Green's function, 
and the superconducting gap function are shown in Fig. 2 
for $t^\prime=0.6$ ($t=1.2$) with $T=0.06$. 
Ridges in the Green's function delineate the Fermi surface.  
Fermi surface comprises two bands (second and third from the bottom), 
which we call band A and band B.  We can see that each band 
has a squre-like pocket for the Fermi surface, 
which is reminiscent of the Fermi surface of 
Sr$_2$RuO$_4$, with a good reason as we shall see.  
On the other hand we notice that the spin susceptibility 
has broad peaks around ($\pi,\pi$).  
Quite unexpectedly, the gap function has a strange 
structure in such as way that its amplitude on the 
Fermi surface is peaked at the {\it corners} rather 
than along the edges.  
\begin{figure}
\begin{center}
\leavevmode
\epsfxsize=5cm
\epsfbox{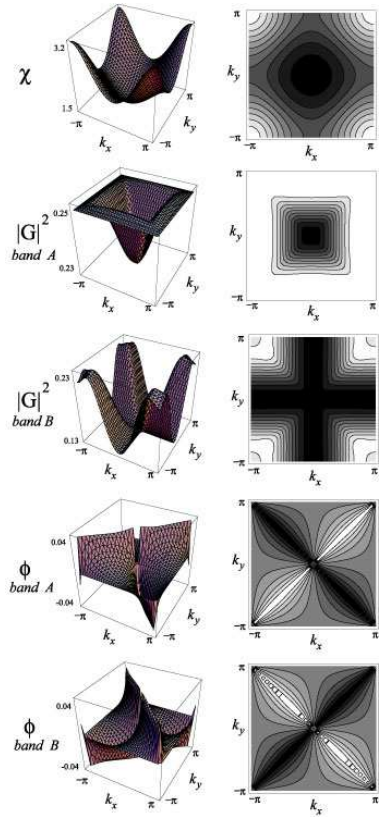}
\vspace*{0.3cm}
\caption{
Spin susceptibility ($\chi$), the 
absolute value of the Green's function squared ($|G|^2$), 
and the superconducting gap function $\Phi$ 
against $k_x, k_y$ for the lowest Matsubara frequency 
for $U=7$, $n=0.85$, $t=1.2$, $t^\prime=0.6$, and $T=0.06$. 
}
\label{fig2}
\end{center}
\end{figure}
Figure 3 shows $T_c$ as a function of $t$.\cite{note}
A plaquette lattice is characterized by the intra-plaquette 
transfer $t$ and inter-plaquette one $t'$.  
To facilitate comparison, we have fixed 
the single-particle bandwidth 
$W=2(2|t|+|t^\prime|)$ to $6$, which is the value of $W$ when 
$t=t'=1$. 
Interestingly, $T_c$ is larger for smaller $|t^\prime|$, 
which is in accord with our expectation that 
the case of plaquette lattice (tetramirization in the 
present case) is favorable.  $T_c$ saturates to a value 
$T_c=0.06=0.01W$, which is two times greater 
than that ($<0.03t$) for the square lattice.  

The enhancement of $T_c$ in smaller $|t^\prime|$ can be
understood by the nesting between the disconnected 
Fermi surface (Fig. 4). 
At half filling ($n=1$) the Fermi surface is perfectly nested 
regardless of the value of $|t^\prime|$. 
When less than half-filled,     
the nesting in the band depicted by $\Vec{Q}$ in Fig. 3 
degrades but not so much for smaller $|t^\prime|$. 
Because the effective attraction here is mediated by the 
antiferromagnetic spin fluctuation (with a large $|Q|$), 
$T_c$ is larger for better nesting, 
i.e., for smaller $|t^\prime| \propto$ the
warping of the quasi-1D Fermi surface. 
This $t^\prime$-dependence of $T_c$ can be understood also in  
real space, where the intra-cell spin singlet 
should be more robustly formed for smaller $|t^\prime|$.

\begin{figure}
\begin{center}
\leavevmode
\epsfxsize=7cm
\epsfbox{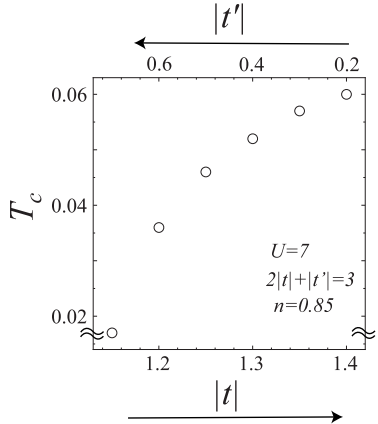}\\
\caption{ 
$T_c$ 
plotted as a function of $t$ for $U=7$ and $n=0.85$ 
in units where the band width $2(2t+t^\prime)=6$.
}
\label{fig3}
\end{center}
\end{figure}
\begin{figure}
\begin{center}
\leavevmode
\epsfxsize=7cm
\epsfbox{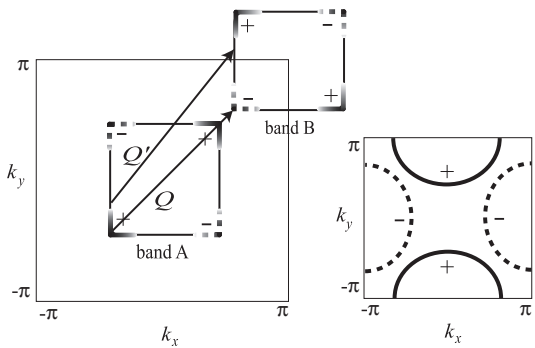}\\
\vspace*{0.5cm}
\caption{
Fermi surfaces of the lattice depicted in Fig.1.  
The solid (dashed) curves represent positive (negative) values
of the gap function on the Fermi surface.
The arrows represent typical momentum transfer 
in pair scattering processes, where $\Vec{Q}$ corresponds to 
the wave vector at which the spin susceptibility is peaked. 
The solid (dashed) curves in the inset represents 
the positive (negative) values of the gap function 
in a plaquette lattice in Ref. {\protect\cite{NTT}}. 
}
\label{fig4}
\end{center}
\end{figure}
We can explain the structures of the strange gap function 
in terms of molecular orbitals (Fig. 5a) 
and the pairing (Fig. 5b) in real space. 
As mentioned, the Fermi surface reminds us of the ($\alpha$ 
and $\beta$) bands in SrRuO, which consists of two 
sets of quasi-1D array of orbitals.   
In the present case the quasi-1D band structure arises 
from the molecular orbitals in the plaquette.  
The molecular orbitals that are relevant to bands A and B 
have $p_x$- and $p_y$- symmetries (Fig. 5a), 
so that we do have two sets of 1D arrays of orbitals. 
Due to the hybridization of the two orbtials, the 
two sets of quasi-1D Fermi surface (two, orthogonal sets of 
parallel lines) anticross, and we end up with two square-shaped 
pockets. 

Now the interesting point is, the Cooper pair, 
arising from antiferromagnetic fluctuations, 
is formed across the adjacent sites (ellipses in Fig. 5b), 
that is {\it across} the two molecular orbitals.  
This implies that the gap function 
on the Fermi surface is large only around the 
positions at which the two ($p_x$ and $p_y$) bands 
hybridize, i.e., around the anticrossing points. 
The pairings depicted in Fig. 5b as solid ellipses
and dashed ellipses have opposite signs, 
so that the symmetry is $d_{xy}$.  
We have thus a curious case of an inter-band pairing 
(a pairing formed by inter-band pair scatterings).  
We can in fact confirm that the pairing is interband and $d_{xy}$ 
by looking at the sign of the gap function.  
The ellipses is Fig. 5 are represented by 
$\langle C_{A(B)\uparrow}C_{A(B)\downarrow}\rangle$, 
where $C_{A(B)\sigma}\equiv C_{1\sigma}-C_{2(4)\sigma}
+C_{4(2)\sigma}-C_{3\sigma}$, where 
$C_{i\sigma}$ annihilates an electron 
with spin $\sigma$ at site $i$ as numbered in Fig.5. 
When the pairing is formed across adjacent sites
we can show $\langle C_{A\uparrow}C_{A\downarrow}\rangle=- 
\langle C_{B\uparrow}C_{B\downarrow}\rangle$ (opposite 
signs across the bands) with an odd parity ($d_{xy}$) against 
$k_x \leftrightarrow -k_x$ or $k_y \leftrightarrow -k_y$.  

This explains why the gap function 
changes sign within each pocket.  
Hence the $s$-wave symmetry on each pocket is 
not a necessary condition for a high $T_c$. 
Even when there are intra-pocket sign changes, 
the disconnectivity of the Fermi surface 
plays an important role, since the {\it inter-pocket} 
sign change associated with $\Vec{Q}$ 
does exploit this in avoiding intersecting the Fermi surface. 

One might then wonder why the intra-pocket sign changes do not 
reduce $T_c$, since the pair-scattering 
processes depicted by $\Vec{Q}^\prime$ in Fig. 4 
connecting the $k$-points across which 
the gap function has the same sign 
would normally reduce $T_c$.   
In the present case, 
however, we have an ingeneous situation where 
the gap function is strongly peaked at the 
position (corners in this example) where hybridization occurs, 
so that the 
$\Vec{Q}^\prime$ processes, being off the peak, have little effect.  

How is the present model compared with 
the lattice considered in Ref. \cite{NTT}, 
where the plaquette is a four-membered ring in 
either case.  Here the consideration in real space 
comes in handy.  
The latter case is a 2D system rather than quasi-1D, 
where the pockets arise due to a band folding 
due to the tetramerization,.  
The pairing in real space 
is formed within each molecular orbital (Fig. 6) unlike 
the present case, 
so the superconducting gap function has an $s$-wave symmetry 
on each Fermi surface pocket (but opposite signs 
across the two pockets with the nesting $\Vec{Q} \approx 
0$ due to the band folding (inset of Fig. 4). 
Despite of this distinction, the resulting 
$T_c$'s are similar between the two cases.  

Another interesting comparison is 
with Sr$_2$RuO$_4$ \cite{SRO,KOAA}, 
which also suggests an importance of the 
real space structure. 
Although the Fermi surface is quasi-1D in Sr$_2$RuO$_4$ as well, 
the oxide has strong pieces of evidence for triplet pairing, 
and even when one considers the spin-singlet pairing, 
$T_c$ (for the $\alpha, \beta$ bands for this material) estimated 
with the FLEX is very small\cite{KOAA}. 
Due to the quasi-one dimensionality 
the spin susceptibility $\chi(\Vec{q},\omega)$ has linear ridges 
in k-space.  
As a result, the pair scatterings have 
large contributions all over the ridges, 
so that the (extended) $s$-wave gap function involves 
unfavorable pair scatterings across which the gap has 
the same sign, resulting in the reduced $T_c$. 
This makes the spin-triplet, $p$-wave pairing 
more favorable, for which the resulting gap function has 
large amplitudes all over the Fermi surface, except for 
dips at the corners . 
By contrast, the present case has a built-in 
antiferromagnetic structure within the unit cell in real space, 
so that the spin susceptibility has a 
well-defined peak around ($\pi,\pi$). 
This makes a specific $\Vec{Q}$ to be relevant 
in the pair scattering, which makes 
the gap function peaked at specific points, which in turn 
gives rise to an enhancement in $T_c$. 

To summarize, we should question not only 
the structure of the Fermi surface but also 
the structure of the molecular orbitals and the superconducting pairings 
in real space in understanding the superconductivity 
from the electron repulsion. 
One interesting tendency is that 
high $T_c$'s are found in plaquette systems where unit cells 
having multiple sites are connected 
with a relatively small inter-plaquette 
hopping \cite{Kuroki&Arita,NTT,KTA}, 
which also suggests the importance of the real space picture.  
In fact, the relation between real-space and the momentum-space
pictures has been discussed for systems consisting of dimers 
recently.\cite{KTA}
It is an interesting future problem to see how the real-space 
and the momentum-space pictures are related with each other 
for wider class of systems in realizing high $T_c$'s. 

We wish to thank Professor J. Akimitsu for valuable discussions. 
This work was supported in part by a Grant-in-Aid for Scientific 
Research and Special Coordination Funds from the Ministry of 
Education of Japan. 
Numerical calculations were performed at the Supercomputer Center,
Institute for Solid State Physics, University of Tokyo.
\begin{figure}
\begin{center}
\leavevmode
\epsfxsize=7cm
\epsfbox{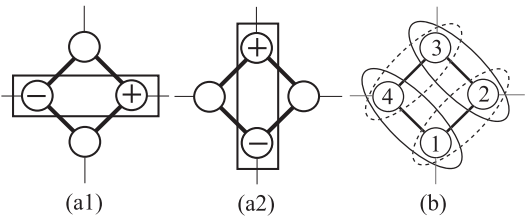}\\
\vspace*{0.5cm}
\caption{
(a) Intra-plaquette molecular orbitals with $p_x$ (a1) and 
$p_y$ (a2) symmetries.  (b) Cooper pairs 
with solid and dashed ellipses indicating  
pairing amplitudes with opposite signs. 
}
\label{fig5}
\end{center}
\end{figure}

\begin{figure}
\begin{center}
\leavevmode
\epsfxsize=7cm
\epsfbox{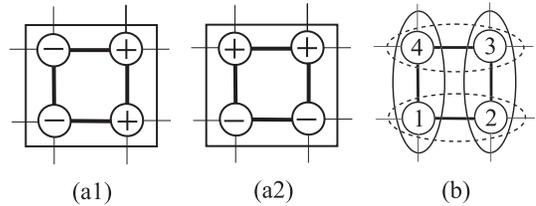}\\
\vspace*{0.5cm}
\caption{
(a) Intra-plaquette molecular orbitals with $p_x$(a1) and 
$p_y$(a2) symmetries, and  (b) Cooper pairs 
for the lattice considered in Ref.\protect\cite{NTT}.
}
\label{fig6}
\end{center}
\end{figure}

\end{multicols}

\begin{references}
\bibitem{note} 
$T_c$  depends only on the absolute values of $t$ or $t^\prime$. 
The invariance under $t\rightarrow-t$ and/or 
$t^\prime\rightarrow -t^\prime$ can be shown with a 
gauge transformation, $C_{1(3)}\rightarrow-C_{1(3)}$ 
and $C_{2(4)}\rightarrow-C_{2(4)}$ for 
$t\rightarrow -t$, and $C_n\rightarrow(-1)^{x+y}C_n$ ($n=1,2,3,4$) 
for $t^\prime\rightarrow-t^\prime$, 
where $x$ and $y$ are the position of the unit cell measured in units of 
the cell size.
\end{references}
\end{document}